\documentstyle[12pt,a4wide,epsf]{article}
\newcommand{\nin}{\noindent}
\newcommand{\be}{\begin{equation}}
\newcommand{\ee}{\end{equation}}
\newcommand{\bea}{\begin{eqnarray}}
\newcommand{\eea}{\end{eqnarray}}
\newcommand{\br}{\hskip .25cm/\hskip -.25cm}

\newcommand{\nn}{\nonumber\\}

\newcommand{\ol}{\overline}

\begin{document}

KCL-PH-TH/2011-{\bf 07}

\begin{center}

{\Large{\bf On higher-order corrections\\ in a four-fermion Lifshitz model}}

\vspace{1cm}

{\bf J.~Alexandre}\footnote{jean.alexandre@kcl.ac.uk}, {\bf J.~Brister}\footnote{james.brister@kcl.ac.uk} and 
{\bf N.~Houston}\footnote{nicholas.houston@kcl.ac.uk}\\
King's College London, Department of Physics, London WC2R 2LS, UK

\vspace{1cm}

{\bf Abstract}

\end{center}

\vspace{0.5cm}

\nin We study a flavour-violating four-fermion interaction in the Lifshitz context, in 3+1 dimensions and with a critical exponent
$z=3$. This model is renormalizable, and features
dynamical mass generation, as well as asymptotic freedom. At one-loop, it is only logarithmically divergent, but 
the superficial degree of divergence of the two-point functions is 3. We calculate the two-loop corrections 
to the propagators, and show that, at this order, the Lorentz-violating corrections to the IR dispersion relation are quadratic
in the cut off. Furthermore, these corrections are too important to represent a physical effect.
As a consequence, 
the predictive power of the model in terms of Lorentz-violating effects in the propagation of particles is limited.

\vspace{1.5cm}

\section{Introduction}

Lifshitz-type theories, where time and space have different mass dimensions and therefore violate Lorentz invariance, 
have attracted attention in recent years, motivated essentially by the possibility of defining new renormalizable interactions. 
This is because of an improvement in the convergence of loop integrals, due to higher order space derivatives, which is achieved 
without the introduction of ghost degrees of freedom, since the order of time derivatives remains minimal. 
A review of Lifshitz theories can be found in \cite{reviewLifshitz}
for quantum field theories in particle physics and in \cite{reviewHorava} for the Horava-Lifshitz alternative approach to gravity. 
An example of a new renormalizable interaction in the Lifshitz context is the Liouville interaction in 3+1 dimensions \cite{LL}, 
where the exponential potential for the scalar field
is a relevant interaction because the field is dimensionless if the critical exponent is $z=3$.

We consider a Lifshitz-type four-fermion interaction model, which,
in $d=3$ space dimensions and for an anisotropic scaling $z=3$, is renormalizable \cite{Anselmi1}.
Such models have been studied in \cite{4fermion}, where two fundamental 
properties were shown: dynamical mass generation and asymptotic freedom. One motivation for these theories, besides renormalizability, 
is the apparent improvement of quantum corrections, since the one-loop graphs are only logarithmically divergent, instead of quadratically
in the Lorentz case. But the overall superficial degree of divergence of the graphs of the theory is actually $\omega=6-3E/2$,
where $E$ is the number of external lines. If one considers the propagator ($E=2$), the corresponding corrections have 
a superficial degree of divergence equal to 3.
The coefficient of the cubic divergence may cancel for some graphs, but we calculate here
a two-loop graph which shows that the divergence in the model is at least quadratic. Therefore, although 
renormalizable, this Lifshitz model still contains ``large'' divergences.

Our model features two massless fermion flavours, coupled with four-fermion interactions which do not respect flavour symmetry.
After showing the occurrence of dynamical flavour oscillations in this model, 
we calculate the modified dispersion relations for these two fermions, arising from quantum fluctuations.
Classically, all fermions have the same dispersion relations, with higher order powers of the space momentum $\vec p$, rescaled by  
a large mass $M$, which represents the crossover scale between the Lifshitz and Lorentz regimes. 
These dispersion relations coincide with the expected Lorentz-invariant one in the infrared (IR) regime $|\vec p|<<M$.
Taking into account quantum corrections, though, modifies this IR limit: it is known in Lifshitz-type studies that different species of particles
see different effective light cones \cite{IengoRussoSerone}. 
Since our model breaks flavour symmetry, the dressed IR dispersion relations are different from 
the Lorentz-invariant one, and we show that the corresponding corrections are quadratically divergent. Furthermore, as we will see, 
these corrections are too important to 
represent any physical effect. As a consequence, a proper treatment of the 
model would consist in defining counterterms to absorb these divergences, and no prediction can be made as far as Lorentz-violating 
propagation is concerned 
(a logarithmic divergence could lead to a ``realistic'' energy dependent effective maximmum speed). 

The next section derives the dynamical masses for the model, including the mass mixing terms necessary for flavour oscillations.
From our study of dynamically induced flavour oscillations, and taking into account experimental data on neutrino oscillations, 
we derive values for the coupling constants of our model, which, as expected, are perturbative. Although neutrinos are not Dirac fermions, 
the corresponding experimental constraints give a good order of magnitude for the parameters in our model. 
Section 3 shows the asymptotic freedom of the interaction, 
based on a one-loop calculation. The four-point function has a vanishing superficial degree of divergence, such that higher order corrections
cannot change the sign of these beta functions. The effective IR dispersion relations, dressed by quantum fluctuations, are derived in section 4. 
For this, we need to go to two loops, since the one-loop correction to the fermion propagators is momentum-independent. 
Detailed calculations are given in the Appendix, where we perform part of the integration analytically and then integrate the rest numerically.

\section{Flavour oscillations}

We describe here, in the Lifshitz context, how flavour oscillations can arise dynamically from flavour-mixing interactions
between two massless bare fermions, as suggested in \cite{Anselmi2}. 
From our expressions for the dynamical masses, together with experimental data, we find phenomenologically realistic values for the coupling 
constants of our model.\\
Oscillations of massless neutrinos are studied in \cite{Benatti}, where neutrinos are considered open systems, interacting with an
environment. Such oscillations have also been studied in \cite{masslessLV}, in the framework of Lorentz-violating models, 
involving non-vanishing vacuum expectation values for vectors and tensors. Whilst these studies have been questioned by phenomenological 
constraints \cite{excluded}, our present model, based on anisotropic
space time and higher order space derivatives, is not excluded.\\
Flavour oscillations were also related to superluminality in \cite{Magueijo}, where it is shown that,
if superluminality is due to a tachyonic mode, the latter can be stabilised by flavour mixing.
Finally, in \cite{Morris}, superluminal effects are related to the extension of a single neutrino wave function, where the
oscillation mechanism plays a role in the uncertainty of the neutrino position.

\subsection{Flavour symmetry violating 4-fermion interactions}

We work in the $z=3$ Lifshitz context, in $d=3$ space dimensions.  
We consider two flavours of massless Dirac fermions $\psi_1,\psi_2$, and the free action
\be
S_{free}=\int dt d\vec x \Bigg( \ol\psi_a i\gamma^0\dot\psi_a-\ol\psi_a(M^2-\Delta)(i\vec\partial\cdot\vec\gamma)\psi_a\Bigg)
~~~~~~a=1,2~,
\ee
where $[M]=1$ and $[\psi_a]=3/2$, and a dot over a field represents a time derivative. 
For the dispersion relations to be consistent in the IR (see eq.(\ref{disprel}) below), one can consider $M$ typically of
the order of a Grand Unified Theory scale (GUT), although we will show that our results only slightly depend on the actual value of $M$.
We introduce the following renormalizable, flavour-violating and attractive 4-fermion interactions 
\be\label{Sint}
S_{int}=\int dtd\vec x\Big(g_1\ol\psi_1  \psi_1+g_2\ol\psi_2  \psi_2+h(\ol\psi_1  \psi_2+\ol\psi_2  \psi_1)\Big)^2~,
\ee
where the coupling constants $g_1,g_2,h$ are dimensionless.
As shown in \cite{4fermion}, this kind of model exhibits dynamical mass generation, which can be seen only with a non-perturbative 
approach, as will be shown in the next section.
Taking into account the dynamical masses, but ignoring quantum corrections to the kinetic terms, the dispersion relations are
of the form
\be
\omega^2=m_{dyn}^6+(M^2+p^2)^2p^2~,~~~~~a=1,2~,
\ee
which, after the rescaling $\omega=M^2\tilde\omega$, leads to
\be\label{disprel}
\tilde\omega^2=\tilde m_{dyn}^2+p^2+\frac{2p^4}{M^2}+\frac{p^6}{M^4}~,
\ee
where $\tilde m_{dyn}=m_{dyn}^3/M^2$. One can see then that Lorentz-like kinematics are recovered in the IR regime $p^2<<M^2$, as
expected in the framework of Lifshitz models. After the rescaling $t=\tilde t/M^2$, the action reads
\bea
S&=&\int d\tilde t d\vec x \Bigg( \ol\psi_a i\br\partial\psi_a+\ol\psi_a\frac{\Delta}{M^2}(i\vec\partial\cdot\vec\gamma)\psi_a\nn
&&+\left[\frac{g_1}{M}\ol\psi_1  \psi_1+\frac{g_2}{M}\ol\psi_2  \psi_2+\frac{h}{M}(\ol\psi_1  \psi_2+\ol\psi_2  \psi_1)\right]^2\Bigg)~,
\eea
where we can see that the four fermion couplings $(g_a/M)^2,~g_ah/M^2$ and $(h/M)^2$ are very small compared to the Fermi coupling 
$\simeq 10^{-5}$ GeV$^{-2}$, if $M$ is of the order of a GUT scale, or even several orders of magnitude smaller, and $g_a,h$ are perturbative.
Finally, note that, for Large Hadron Collider energies up to few TeVs,  
the classical Lifshitz corrections $p^4/M^2$ and $p^6/M^4$ in the dispersion relation (\ref{disprel}) are not detectable,
if $M$ is of the order of a GUT scale. 
For this reason, if one wishes to describe measurable non-relativistic effects in the Lifshitz context, 
these should be sought in quantum corrections to the IR dispersion relation.

\subsection{Superficial degree of divergence}

It is interesting to note that, although this Lifshitz model has only logarithmic divergences at one-loop,
quantum corrections actually do not ``behave better'' than those in the Lorentz-invariant $\phi^4$ theory, since  
the superficial degree of divergence of the propagator is 3.\\
To show this, we calculate via the usual approach the degree of divergence $\omega$ of a graph with $E$ external lines.
Each loop gives an integration measure $dp_0d^3p$, which has mass dimension 6, and each propagator has mass dimension -3.
For a graph with $I$ internal lines and $L$ loops, the superficial degree of divergence is therefore
$\omega=6L-3I$. As usual, because of momentum conservation, we also have $L=I-n+1$, 
where $n$ is the number of vertices of the graph. Finally, since we have 4-leg vertices, we also have the relation $4n=E+2I$. 
Taking into account these constraints, we find $\omega=6-3E/2$.\\
From this result, we see that the four-point function is at most logarithmically divergent, but the propagator has
a superficial degree of divergence equal to 3, although the one-loop mass corrections are logarithmically divergent only, 
as we show in the next subsection.

\subsection{Dynamical generation of masses}

We now calculate the dynamical masses generated by the interaction (\ref{Sint}).
For this, we introduce the auxiliary scalar field $\phi$ to express the interaction as
\bea
\exp\left( iS_{int}\right) &=&\int{\cal D}[\phi]\exp(iS_\phi)~, \nn
\mbox{with}~~~~~~~~~~~S_\phi&=&\int dtd\vec x\left(-\phi^2+2\phi(g_1\ol\psi_1  \psi_1+g_2\ol\psi_2  \psi_2+h(\ol\psi_1  \psi_2+\ol\psi_2  \psi_1)
\right)~,
\eea
and then calculate the effective potential for $\phi$=constant as
\be
\exp\left( i{\cal V}V_{eff}(\phi)\right)=\int{\cal D}[\psi_1,\ol\psi_1  ,\psi_2,\ol\psi_2  ]\exp\left( iS_{free}+iS_\phi\right) ~,
\ee
where ${\cal V}$ is the space time volume.
This integration can be done exactly, since $S_{free}+S_\phi$ is quadratic in fermion fields, and leads to an effective potential for $\phi$.
From the dispersion relation (\ref{disprel}), one can see that
a non-trivial minimum $\phi_{min}$ for this effective potential will give the flavour mixing mass matrix 
\be
\left(\begin{array}{cc}m_1^3&\mu^3\\\mu^3&m_2^3\end{array}\right)=2\phi_{min}
\left(\begin{array}{cc}g_1&h\\h&g_2\end{array}\right)~,
\ee
which leads to the rescaled masses
\be\label{rescaledm}
\left(\begin{array}{cc}\tilde m_1&\tilde\mu\\\tilde\mu&\tilde m_2\end{array}\right)=2\frac{\phi_{min}}{M^2}
\left(\begin{array}{cc}g_1&h\\h&g_2\end{array}\right)~.
\ee
As a consequence, the mass eigenstates are
\be\label{eigenstates}
m_\pm=\frac{\phi_{min}}{M^2}\left( g_1+g_2\pm(g_1-g_2)\sqrt{1+\tan^2(2\theta)}\right) ~,
\ee
where the mixing angle $\theta$ is defined by
\be\label{tan2theta}
\tan(2\theta)\equiv\frac{2h}{g_1-g_2}~.
\ee
With the auxiliary field, the Lagrangian can then be written in the form $\ol\Psi{\cal O}\Psi$,
where 
\be
\Psi=\left(\begin{array}{c}\psi_1\\ \psi_2\end{array}\right)~,
\ee 
and the operator ${\cal O}$ is
\be
{\cal O}=\left(\begin{array}{cc}i\gamma^0\partial_0-(M^2-\Delta)(i\vec\partial\cdot\vec\gamma)+2g_1\phi&2h\phi\\
2h\phi&i\gamma^0\partial_0-(M^2-\Delta)(i\vec\partial\cdot\vec\gamma)+2g_2\phi\end{array}\right)~.
\ee
Integration over the fermions then gives the following effective potential for $\phi$ (where the Euclidean metric is used for the loop momentum)
\bea
V_{eff}(\phi)&=&\phi^2-\frac{1}{2}\int\frac{d\omega}{2\pi}\frac{d\vec p}{(2\pi)^3}
\ln\Bigg(\Big[\omega^2+(M^2+p^2)^2p^2\Big]^2\\
&&~~~~~~~~~~~~~~~+4\phi^2\Big[\omega^2+(M^2+p^2)^2p^2\Big](g_1^2+g_2^2+2h^2)+16(g_1g_2-h^2)^2\phi^4\Bigg) ~.\nonumber
\eea
A derivative with respect to $\phi$ gives
\be
\frac{dV_{eff}}{d\phi}=2\phi-\phi\int\frac{d\omega}{2\pi}\frac{d\vec p}{(2\pi)^3}
\frac{A\omega^2+B}{(\omega^2+C_+)(\omega^2+C_-)}~,
\ee
where 
\bea
A&=&4(g_1^2+g_2^2+2h^2)\\
B&=&4(M^2+p^2)^2p^2(g_1^2+g_2^2+2h^2)+32\phi^2(g_1g_2-h^2)^2\nn
C_\pm&=&(M^2+p^2)^2p^2+2\phi^2\left[ (g_1^2+g_2^2+2h^2)\pm\sqrt{(g_1^2-g_2^2)^2+4h^2(g_1+g_2)^2}\right] ~.\nn
\eea
The integration over frequencies $\omega$ leads to
\be
\frac{dV_{eff}}{d\phi}=2\phi-\phi\frac{1}{(2\pi)^2}\int_0^\Lambda p^2dp\left( \frac{B}{C_+\sqrt C_-+C_-\sqrt C_+}
+\frac{A}{\sqrt C_++\sqrt C_-}\right) 
\ee
where $\Lambda$ is the UV cut off, assumed to be large compared to $M$. A non-trivial minimum $\phi_{min}\ne0$ 
for this effective potential is solution of the equation
\be\label{gap}
8\pi^2=\int_0^\Lambda p^2dp\left( \frac{B}{C_+\sqrt C_-+C_-\sqrt C_+}+\frac{A}{\sqrt C_++\sqrt C_-}\right)~.
\ee
The dominant contribution of these logarithmically divergent integrals comes form $p\to\Lambda$, 
so we can therefore approximate 
\be
C_-\simeq C_+\simeq p^6+\frac{A}{2}\phi^2~~~~\mbox{and}~~~~B\simeq Ap^6~,
\ee
such that
\be
\int_0^\Lambda p^2dp\left( \frac{B}{C_+\sqrt C_-+C_-\sqrt C_+}+\frac{A}{\sqrt C_++\sqrt C_-}\right)
\simeq \frac{A}{3}\ln\left( \frac{2\sqrt 2\Lambda^3}{\phi\sqrt{A}}\right) ~.
\ee
The gap equation (\ref{gap}) then gives
\be
\phi_{min}\simeq\frac{\Lambda^3}{\sqrt{g_1^2+g_2^2+2h^2}}\exp\left( \frac{-6\pi^2}{g_1^2+g_2^2+2h^2}\right)~,
\ee
and the rescaled masses (\ref{rescaledm}) are 
\bea\label{finalmasses}
\tilde m_a&\simeq&\frac{2g_a}{\sqrt{g_1^2+g_2^2+2h^2}}~\frac{\Lambda^3}{M^2}\exp\left( \frac{-6\pi^2}{g_1^2+g_2^2+2h^2}\right)\\
\tilde\mu&\simeq&\frac{2h}{\sqrt{g_1^2+g_2^2+2h^2}}~\frac{\Lambda^3}{M^2}\exp\left( \frac{-6\pi^2}{g_1^2+g_2^2+2h^2}\right)~.
\eea
As expected, these masses are not analytical in the coupling constants and could not have been obtained with a perturbative expansion.
Similar results have been obtained in the context of magnetic catalysis \cite{mag}, based on the Schwinger-Dyson approach, and also
for Lorentz-violating extensions of $QED$ \cite{LVQED}. Neither of these studies however feature anisotropic space time studied herein.\\ 
Finally, we note that the approach 
adopted here, based on the effective potential for the auxiliary field $\phi$, is in principle valid for a large number of flavours. 
Indeed, this auxiliary field depends on space and time, and its fluctuations around the minimum $\phi_{min}$ induce new fermion 
interactions. For $N$ fermion flavours though, these fluctuations are suppressed by $1/N$, which justifies the approach. In our case, $N=2$
is not ``large'', but the corresponding order of magnitude for the dynamical masses is sufficient for a suitably accurate determination
of the coupling constants $g_1,g_2$, as explained in the next subsection.

\subsection{Experimental constraints}

\begin{figure}
\epsfxsize=8cm
\centerline{\epsffile{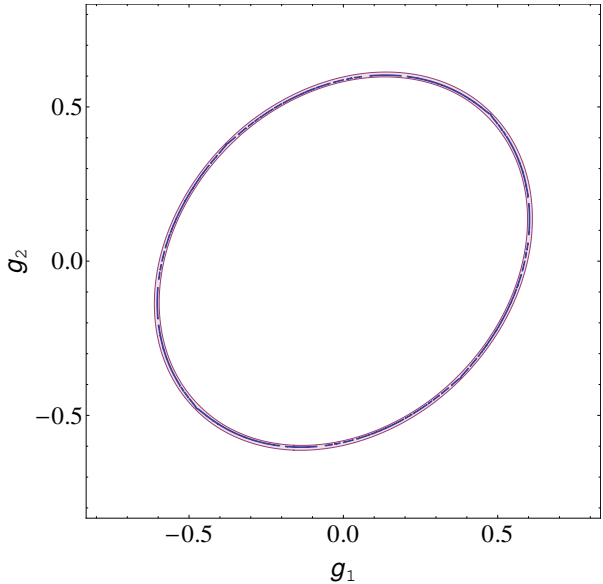}}
\caption{Values of $g_1$ (x-axis) and $g_2$ (y-axis) allowed by experimental constraints, for $M/\Lambda=10^{-11}$. Negative 
values are allowed, since the physical quantities depend on the square of the coupling constants only. Points where $g_1=g_2$ are
strictly speaking not allowed, since at these points $\Delta m^2=0$. However, the resulting logarithmic singularity is very localised
in the parameter space, such that we can safely chose $g_1$ and $g_2$ perturbatively close to each other.}
\label{g1g2}
\end{figure}

From the expressions (\ref{eigenstates}), we obtain the following difference of mass eigenstates squared 
\bea\label{deltam2}
\Delta m^2&=&\frac{4}{\cos(2\theta)}\frac{g_1^2-g_2^2}{g_1^2+g_2^2+\tan^2(2\theta)(g_1-g_2)^2/2}\nn
&&~~~~~~~~~\times\frac{\Lambda^6}{M^4}\exp\left( \frac{-12\pi^2}{g_1^2+g_2^2+\tan^2(2\theta)(g_1-g_2)^2/2}\right)~.
\eea
Experimental constraints are \cite{exp}
\bea\label{expconst}
\Delta m^2_{12}&=&7.59(7.22-8.03)\times 10^{-5} ~\mbox{(eV)}^2\nn
\sin^2\theta_{12}&=&0.318(0.29-0.36)~,
\eea
and we plot in fig.(\ref{g1g2}), from the expression (\ref{deltam2}), the set of points in the plane $g_1,g_2$ which are allowed, 
given the experimental constraints (\ref{expconst}). 
We consider $\Lambda\simeq10^{19}$ GeV, corresponding to the Plank mass.
An important property is that the result is hardly sensitive to the value of the mass scales $M$: because 
of the exponential dependence in eq.(\ref{deltam2}), an increase of several orders of magnitude in $M$ leads to an increase of few percent
only for the couplings $g_a$, as shown in the following table. Considering the situation where $h<<1$, such that
$g_1\simeq g_2$, according to eq.(\ref{tan2theta}), the approximate common value for the coupling constants as a function of  
the ratio $M/\Lambda$ is then:

\label{g1=g2}
\begin{table}[h]
\centering
\begin{tabular}{|c||c|c|c|c|c|c|}
\hline
$M/\Lambda$ & $10^{-16}$ & $10^{-15}$ & $10^{-14}$ & $10^{-13}$ & $10^{-12}$ & $10^{-11}$\\ 
\hline
$g_1\simeq g_2$ & 0.46 & 0.47 & 0.48 & 0.48 & 0.49 & 0.50 \\
\hline
\hline
$M/\Lambda$& $10^{-10}$ & $10^{-9}$ & $10^{-8}$ & $10^{-7}$ & $10^{-6}$ & $10^{-5}$\\
\hline
$g_1\simeq g_2$ & 0.51 & 0.53 & 0.54 & 0.55 & 0.56 & 0.58\\
\hline
\end{tabular}
\caption{Coupling constants for different mass scales $M$, when $h<<1$ and $\Lambda=10^{19}$GeV.}
\end{table}

\nin On fig.(\ref{g1g2}), the thin line represents the set of points satisfying the constraint
\be
\left|\ln\left( \frac{\Delta m^2_{experimental}}{\Delta m^2_{calculated}}\right) \right|\le1~,
\ee
and the thick line represents the set of points such that the largest mass eigenvalue is between $10^{-3}$ and 1 eV. We see that
the coupling constants appearing in the theory are then of the order $g_a^2\simeq0.25$, and can be considered perturbative.

\section{Asymptotic freedom}

We now calculate the one-loop coupling constants, for $h<<1$, and we show that the theory is asymptotically free. 
For simplicity, we set $h=0$ but still keep $g_1\ne g_2$.\\
The bare interaction can be expressed as
\be
g_1^2~(\ol\psi_1  \psi_1)^2+G~\ol\psi_1  \psi_1\ol\psi_2  \psi_2+g_2^2~(\ol\psi_2  \psi_2)^2~,~~~~~~~G=2g_1g_2~,
\ee
and the dressed interaction is of the form
\be
(g_1^2+\delta g_1^2)(\ol\psi_1  \psi_1)^2+(G+\delta G)\ol\psi_1  \psi_1\ol\psi_2  \psi_2+(g_2^2+\delta g_2^2)(\ol\psi_2  \psi_2)^2~.
\ee 
Note that no symmetry imposes any relation between $\delta G$ and $\delta g_1^2,\delta g_2^2$: the interaction $\ol\psi_1  \psi_1\ol\psi_2  \psi_2$
is dressed independently of the interactions $(\ol\psi_1  \psi_1)^2$ and $(\ol\psi_2  \psi_2)^2$.

\subsection{One-loop Fermi coupling}

\begin{figure}
\epsfxsize=16cm
\centerline{\epsffile{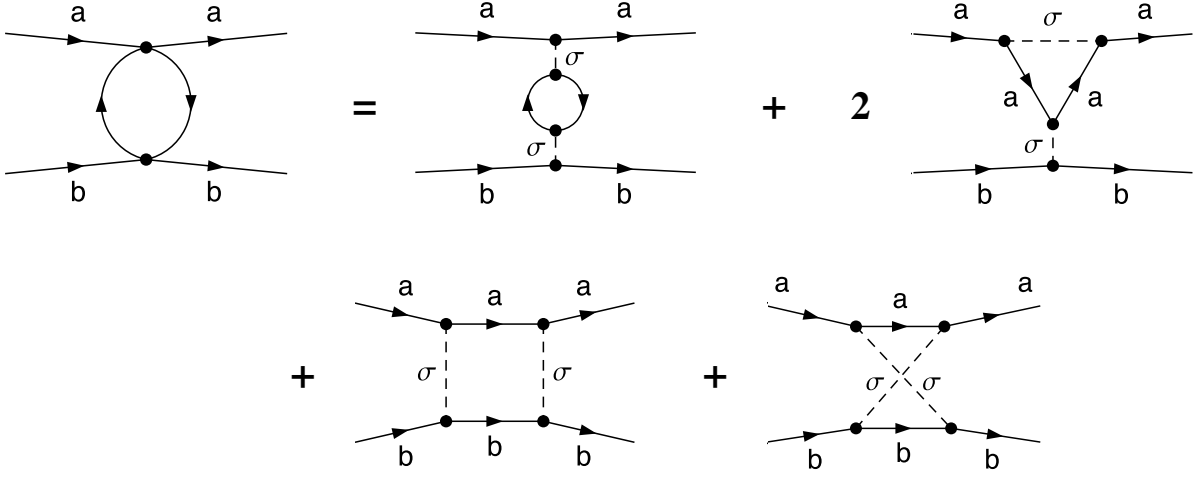}}
\caption{One-loop graphs involving the auxiliary scalar field, which contribute to the four-point function. Solid lines represent
fermions and dashed lines represent the scalar. 
Only the first two diagrams, where the fermion lines cross an odd number of vertices,
contribute to the four-fermion beta functions. 
The first graph corresponds to the insertion of a scalar self-energy, for each flavour, and involves a factor -4 for the trace over Dirac indices.
The second graph has two contributions: one for each insertion of a vertex correction.}
\label{4point}
\end{figure}

If one denotes 
\bea
N_a(\omega,\vec p)&=&\omega\gamma^0-(M^2+p^2)(\vec p\cdot\vec\gamma)+m_a^3\nn
D_a(\omega,\vec p)&=&\omega^2-(M^2+p^2)^2p^2-m_a^6~,
\eea
the generic graph for the one-loop corrections is
\be
I_{ab}=\int\frac{d\omega d\vec p}{(2\pi)^4}~\frac{iN_a(\omega,\vec p)iN_b(\omega,\vec p)}{D_a(\omega,\vec p)D_b(\omega,\vec p)}.
\ee
When $\Lambda>>M$, we obtain
\bea\label{Iab}
I_{ab}&=&-\int\frac{d\omega d\vec p}{(2\pi)^4}\left( \frac{m_a^3/(m_a^3-m_b^3)}{\omega^2-(M^2+p^2)^2p^2-m_a^6}
-\frac{m_b^3/(m_a^3-m_b^3)}{\omega^2-(M^2+p^2)^2p^2-m_b^6}\right)\\
&=&\frac{i}{4\pi^2(m_a^3-m_b^3)}\int_0^\Lambda p^2dp\left( \frac{m_a^3}{\sqrt{(M^2+p^2)^2p^2+m_a^6}}
-\frac{m_b^3}{\sqrt{(M^2+p^2)^2p^2+m_b^6}}\right) \nn
&\simeq&\frac{i}{12\pi^2(m_a^3-m_b^3)}\int_0^{\Lambda^3}dx\left( \frac{m_a^3}{\sqrt{x^2+m_a^3}}-\frac{m_b^3}{\sqrt{x^2+m_b^3}}\right) \nn
&\simeq&\frac{i}{4\pi^2(m_a^3-m_b^3)}\left[ m_a^3\ln\left( \frac{\Lambda}{m_a}\right) -m_b^3\ln\left( \frac{\Lambda}{m_b}\right)\right] ~.\nonumber
\eea
The integral (\ref{Iab}) diverges logarithmically, unlike the Lorentz symmetric case where it diverges quadratically. 
Note that, when $m_b\to m_a$, the previous result is regular and leads to
\be\label{Iaa}
I_{aa}\simeq\frac{i}{4\pi^2}\ln\left( \frac{\Lambda}{m_a}\right)~.
\ee
In order to calculate the number of graphs (\ref{Iab}) contributing to the coupling corrections, 
we introduce the auxiliary field $\sigma$ and write the four-fermion interactions in the form
\be\label{yukawa}
-\frac{1}{2}\sigma^2+\sigma\sqrt2(g_1\ol\psi_1  \psi_1+g_2\ol\psi_2  \psi_2)~.
\ee 
The scalar $\sigma$ does not propagate, but is described by a fictitious propagator, which carries a factor $i$.
This propagator has to be understood in the limit where it shrinks to a point, leading to
the fermion loops given by the expressions (\ref{Iab}). The two vertices corresponding to the effective Yukawa interactions are $i\sqrt2g_1$ 
and $i\sqrt2g_2$.\\
The graphs corresponding to the four-point function are represented in fig.(\ref{4point}), 
in terms of the equivalent Yukawa interaction (\ref{yukawa}),
where the last two graphs do not contribute to the four-fermion beta functions. Indeed, 
the general structure of the four point function is 
\be
\left<0\left|\psi^\dagger\psi\psi^\dagger\psi\right|0\right>=A{\bf 1}\otimes{\bf 1}+B_i{\bf 1}\otimes\sigma^i
+C_{ij}\sigma^i\otimes\sigma^j~,
\ee
where the Dirac indices are omitted and the tensorial product allows for the two in and two out states. 
$A$ is the only quantity contributing to the coupling constant,
since the corresponding term has no Dirac structure.
The four-point function contains one divergence only, which is logarithmic, such that the divergent graphs are obtained only from 
the highest power of momentum in the numerator of propagators, i.e. from $\vec p\cdot\vec\gamma$ and not the mass term.
The last two graphs of fig.(\ref{4point}) contain continuous lines of fermions
with one internal propagator, such that the divergent part is contained in $C_{ij}$ only. $A$ is finite for these two graphs,
and thus does not contribute to the beta function. 
More generally \cite{GrossNeveu}, to any order of the perturbation theory, 
any graph containing a open fermion line, which meets an even number of vertices, 
does not contribute to the beta functions of the model. One needs an even number of internal lines for the product of gamma matrices
(appearing in $\vec p\cdot\vec\gamma$) to give a diverging term with a non-vanishing trace.

\subsection{Beta-functions}

The divergent one-loop correction to the four-fermion interactions are then given by the first two graphs of fig.(\ref{4point}), 
which are as follows.
\begin{itemize}
\item For the flavour preserving four-fermion interaction:\\ 
{\it (i)} Graphs with the insertion of the one-loop scalar self-energy: both flavours contribute to the fermion loop, which induces 
a factor -4 for the trace over Dirac indices. The contribution is then,
\be
-4i^2(i\sqrt2g_a)^4I_{aa}-4i^2(i\sqrt2g_a)^2(i\sqrt2g_b)^2I_{bb}=16g_a^2(g_a^2I_{aa}+g_b^2I_{bb})~.
\ee
{\it(ii)} Graphs with the insertion of the one-loop Yukawa interaction: only the flavour $a$ plays a role, and the contribution is
\be
2i^2(i\sqrt2g_a)^4I_{aa}=-8g_a^4I_{aa}~.
\ee 
The total contribution must be identified with to correction to the bare graph $i(i\sqrt2g_a)^2$, such that 
\be
i\delta g_a^2=-4g_a^2(g_a^2I_{aa}+2g_b^2I_{bb})~,
\ee
and the corresponding beta function is therefore
\be
\beta_a\equiv\Lambda\frac{\partial(\delta g_a^2)}{\delta\Lambda}=-\frac{g_a^2}{\pi^2}(g_a^2+2g_b^2)~.
\ee
\item For the flavour-changing interaction:\\
{\it(i)} Graphs with the insertion of the one-loop scalar self-energy:
\be
16g_ag_b(g_a^2I_{aa}+g_b^2I_{bb})~;
\ee
{\it(ii)} Graphs with the insertion of the one-loop Yukawa interaction:
\be
-4g_bg_a^3I_{aa}-4g_ag_b^3I_{bb}~;
\ee
The total contribution must be identified with the correction to the bare graph $iG$
\be
i\delta G=-12g_ag_b(g_a^2I_{aa}+g_b^2I_{bb})~,
\ee
and the corresponding beta function is therefore
\be
\beta_G=-3\frac{g_ag_b}{\pi^2}(g_a^2+g_b^2)~.
\ee
\end{itemize}
One can conclude from this one-loop analysis that the theory is asymptotically free, since higer orders also diverge at most 
logarithmically, and cannot change the sign of the one-loop beta functions. 
Note that, when $g_1=g_2$, then $\beta_G=2\beta_a$, as expected from the $O(2)$ symmetry.

\section{Two-loop propagator}

Since Lifshitz theories explicitly break Lorentz symmetry, space and time derivatives are dressed differently by quantum corrections. If one 
considers only one particle, or several particles in a given flavour multiplet, frequency and space momentum can always 
be rescaled in such a way that the particles have the usual Lorentz-like IR dispersion relation (after neglecting the higher order powers 
of the space momentum, suppressed by $M$). However if one considers several particles without flavour symmetry, 
then it becomes necessary to perform a 
flavour-independent rescaling of frequency and space momentum, such that different particles see different effective light cones.
This is the case we consider here.

\begin{figure}
\epsfxsize=16cm
\centerline{\epsffile{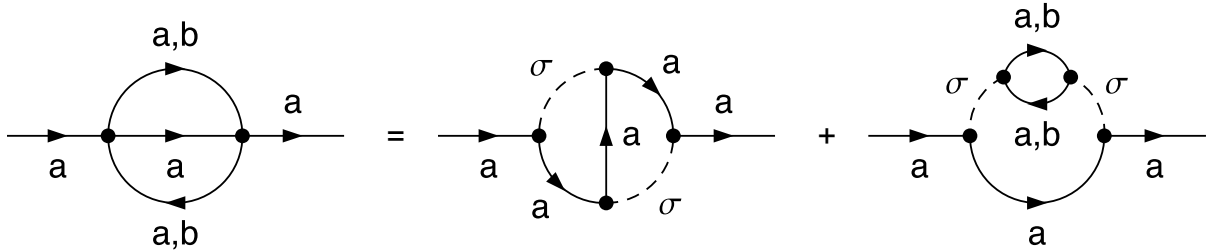}}
\caption{Two-loop contributions to the propagator. The fermion loop in the second graph involves a contribution from each flavour, 
and a factor -4 for the trace over Dirac indices.}
\label{2point}
\end{figure}

\nin The dispersion relations (\ref{disprel}) are not modified at one loop, since the corresponding graphs do not depend on 
the external momentum. We therefore have to go to two loops to find the first quantum corrections, corresponding to the graphs 
represented on fig.(\ref{2point}), in terms of the equivalent Yukawa model (\ref{yukawa}). 

\nin We note here that the two-loop propagator is evaluated in \cite{Gomes^2} for a scalar $\phi^4$ theory, in 6 spacial dimensions and for $z=2$.
This calculation is done in the massless case and in the absence of quadratic space derivatives. Dimensional regularization is used there,
such that the power of the cut off does not appear explicitly in the results. 
The authors conclude that the the Lorentz-symmetry breaking terms flow to 0 in the deep IR.

\subsection{Self energy}

The perturbative graphs on fig. (\ref{2point}) can be calculated with massless bare propagators, 
since the two-loop graphs contain no IR divergence. As a consequence, these graphs are flavour independent (besides an overall 
factor depending on the coupling constants), 
and they involve the integrals
\bea\label{graphs}
I(k_0,\vec k)&=&i^2\int\frac{dp_0d\vec p}{(2\pi)^4}\int\frac{dq_0d\vec q}{(2\pi)^4}~
\frac{iN(-p)iN(-q)iN(p+q+k)}{D(-p)D(-q)D(p+q+k)}\\
J(k_0,\vec k)&=&i^2\int\frac{dp_0d\vec p}{(2\pi)^4}\int\frac{dq_0d\vec q}{(2\pi)^4}~
\frac{\mbox{Tr}[iN(-p)iN(-q)]iN(p+q+k)}{D(-p)D(-q)D(p+q+k)}~,\nonumber
\eea
where the trace in $J$ arises from the fermion loop, and the factors $i^2$ are for the scalar propagators.
Taking into account the different possibilities for the self-energy of flavour $a$, we obtain 
\begin{itemize}
\item $(i\sqrt2g_a)^4I$ for the graph without fermion loop;
\item $[(i\sqrt2g_a)^4+(i\sqrt2g_a)^2(i\sqrt2g_b)^2]J$ for the graph with a fermion loop: one contribution for each flavour in the loop.
\end{itemize}
We calculate these integrals in the Appendix, where we see that the only role of the fermion loop is to give a factor -4 from the
trace over Dirac indices.
We therefore have $J=-4I$, and the total contribution to the momentum-dependent two-loop self energy $\Sigma_a(k_0,\vec k)$ is
given by
\bea\label{sigma}
&&-i\Sigma_a(k_0,\vec k)\\
&=&-4g_a^2(3g_a^2+4g_b^2)I\nn
&=&-i4g_a^2(3g_a^2+4g_b^2)\int\frac{dp_0d\vec p}{(2\pi)^4}\int\frac{dq_0d\vec q}{(2\pi)^4}~
\frac{N(-p)N(-q)N(p+q+k)}{D(-p)D(-q)D(p+q+k)}~.\nonumber
\eea
The bare inverse fermion propagator is  
\be
S_{bare}^{-1}=k_0\gamma^0-M^2\vec k\cdot\vec\gamma+\cdots~,
\ee
where dots represent higher orders in $\vec k$. We parametrize the dressed inverse propagator as
\be
S_{dressed}^{-1}=-m_a^3+(1-Y_a)k_0\gamma^0-(1-Z_a)M^2\vec k\cdot\vec\gamma+\cdots~,
\ee
such that the self energy is
\be\label{self}
\Sigma_a(k_0,\vec k)=S_{bare}^{-1}-S_{dressed}^{-1}=m_a^3+Y_ak_0\gamma^0-Z_aM^2\vec k\cdot\vec\gamma+\cdots
\ee
The integrals (\ref{graphs}) should then be expanded in the external frequency $k_0$ and momentum $\vec k$ in order to find the corrections 
$Y_a,Z_a$. The $k$-independent mass correction $m^3_a$ will be disregarded, since 
the dynamical masses have already been calculated.

\subsection{Dressed dispersion relations}

From the self energy (\ref{self}), the IR dispersion relation for the flavour $a$ is 
\be
(1-Y_a)^2k_0^2= m_a^6+M^4(1-Z_a)^2k^2+\cdots~,
\ee
where $k=|\vec k|$. If we assume that the two fermion flavours are to be coupled to other particles, then one needs a 
flavour-independent rescaling of the dispersion relation. $k_0\to M^2\tilde k_0$ leads then to the following 
product of the phase and the group velocities, $v_p$ and $v_g$ respectively
\be\label{va2}
v_a^2\equiv v_pv_g=\frac{\tilde k_0}{k}\frac{\partial\tilde k_0}{\partial k}=1+2(Y_a-Z_a)+{\cal O}(k/M)^2~.
\ee 
We calculate $Y_a$ and $Z_a$ in the Appendix, by expanding analytically the integral $I$ to first order in $k_0$ and $\vec k\cdot\vec\gamma$, 
and we find a quadratic divergence of the form ($\Lambda>>M$)
\be\label{Y-Z}
Y_a-Z_a\simeq4\kappa ~g_a^2(3g_a^2+4g_b^2)\frac{\Lambda^2}{M^2}~,~~~~\kappa\simeq-3.49\times10^{-5}
\ee
where $a\ne b$. This result shows that the present model is of limited use for the prediction of Lorentz-violating propagation. Indeed,
with the values of $\Lambda/M,g_1,g_2$ shown in Table 1, the result (\ref{Y-Z}) is not perturbative: 
one needs to absorb the quadratic divergence with counterterms, such that the renormalized value of $Y_a-Z_a$ needs to be fixed by 
experimental data. Therefore the model cannot predict quantitative deviations from Special Relativity at low energies.\\
If these corrections were logarithmic, one could infer from our result a cut-off-independent beta function 
for the effective maximum speed seen by the fermions, which could lead to ``realistic'' predictions on potential sub/super-luminal propagation.\\
Note that the rescaling of frequency which leads to the speed squared (\ref{va2}) does not make apparent the fact that, if flavour symmetry is 
exactly satisfied, then the IR dispersion relation are relativistic. If one ignores possible interactions with other particles, 
one can further rescale
\be
k^2=\tilde k^2\frac{1-Y_1}{1-Z_1}\frac{1-Y_2}{1-Z_2}~,
\ee
which leads to the following IR dispersion relations
\bea
\tilde k_0^2&\simeq&\tilde m_1^2+(1+2\delta v)\tilde k^2\nn
\tilde k_0^2&\simeq&\tilde m_2^2+(1-2\delta v)\tilde k^2\nn
\mbox{where}~~\delta v&=&6\kappa(g_1^4-g_2^4)\frac{\Lambda^2}{M^2}~.
\eea
One can see here that the Lorentz-invariant IR dispersion relations are recovered when $g_1=g_2$. 
But for $g_1\ne g_2$, 
one needs the difference $|g_1^2-g_2^2|$ to be proportional to $M^2/\Lambda^2$ in order to deal with realistic phenomenology. 
One can take the example of the largest value $M/\Lambda\simeq10^{-5}$ with $g_a\simeq0.58$ from Table 1.
The upper bound $\delta v\leq2\times10^{-9}$ given by the supernovae SN1987a data \cite{SN1987a} gives then
\be
|g_1^2-g_2^2|\leq\frac{\delta v~M^2/\Lambda^2}{6\kappa(g_1^2+g_2^2)}\simeq10^{-15}~,
\ee
such that flavour symmetry can be considered exact, and the corresponding fine tuning is not natural.

\section{Conclusion}

In the Lifshitz context, we have shown that
flavour oscillations are generated dynamically, for massless fermions coupled via four-fermion interactions.
The corresponding IR dispersion relation is not relativistically invariant, as a consequence of the absence of flavour symmetry.
An essential point is that, even if Lifshitz kinematics are 
hardly detectable at the classical level, quantum corrections lead to important effects in the IR, as flavour oscillations and modified 
dispersion relations.
Nevertheless, we find that modifications to the IR dispersion relations lead to too important Lorentz-violating effects, which are  
obviously ruled out, unless flavour symmetry is almost exactly satisfied, which corresponds to an unnatural fine tuning.\\
We therefore suggest that a realistic Lifshitz model should have logarithmic divergences at most, 
in order to have phenomenological relevance. This is the case, for example, of Lifshitz-type Yukawa models \cite{AFPT}, where 
one-loop corrections to the fermion dispersion relations are finite. Also, Lifshitz-type extensions of gauge theories, 
which are super-renormalizable in 3+1 dimensions and for $z=3$, feature interesting properties \cite{Anselmi3}.
An essential point is, since the gauge coupling constant has a (positive) mass dimension, fermion dynamical mass naturally  
appears in the dressed theory. If, in addition, fermion condensates break flavour symmetry, then vectors automatically become 
massive \cite{massA}, with a mechanism similar to the one initially derived by Schwinger in 1+1 dimensional Quantum Electrodynamics, 
and used in Technicolor studies. Such models are to be looked at in future publications,
in order to determine their phenomenological relevance.

\vspace{0.5cm}

\nin{\bf Acknowledgements} We would like to thank Antonis Tsapalis for useful discussions on the four-fermion model, and Alessio Comisso
for suggestions on numerical integrations.

\section*{Appendix: two-loop propagator}

An expansion in the frequency $k_0$ of the integrand appearing on the right-hand side of eq.(\ref{sigma}) gives 
\bea\label{freqexp}
&&\frac{N(-p_0,-\vec p)N(-q_0,-\vec q)N(p_0+q_0+k_0,\vec p+\vec q)}{D(-p_0,-\vec p)D(-q_0,-\vec q)D(p_0+q_0+k_0,\vec p+\vec q)}
=\frac{N(-p)N(-q)N(p+q)}{D(-p)D(-q)D(p+q)}\\
&&~~~~~+k_0\gamma^0\frac{N(-p)N(-q)}{D(-p)D(-q)D(p+q)}
-2k_0(p_0+q_0)\frac{N(-p)N(-q)N(p+q)}{D(-p)D(-q)D^2(p+q)}+{\cal O}(k_0^2)\nonumber~,
\eea
where $(p)\equiv(p_0,\vec p)$, and an expansion in the spatial momentum $\vec k$ gives
\bea\label{momexp}
&&\frac{N(-p_0,-\vec p)N(-q_0,-\vec q)N(p_0+q_0,\vec p+\vec q+\vec k)}{D(-p_0,-\vec p)D(-q_0,-\vec q)D(p_0+q_0,\vec p+\vec q+\vec k)}
=\frac{N(-p)N(-q)N(p+q)}{D(-p)D(-q)D(p+q)}\\
&&~~~~~-\frac{N(-p)N(-q)}{D(-p)D(-q)D(p+q)}\left[2((\vec p+\vec q)\cdot\vec k)((\vec p+\vec q)\cdot\vec\gamma)
+(M^2+(\vec p+\vec q)^2)(\vec k\cdot\vec\gamma) \right] \nn
&&~~~~~+\frac{N(-p)N(-q)N(p+q)}{D(-p)D(-q)D^2(p+q)}2((\vec p+\vec q)\cdot\vec k)\left[ M^2+(\vec p+\vec q)^2\right]
\left[ M^2+3(\vec p+\vec q)^2\right] +{\cal O}(k_0^2)\nonumber~.
\eea

\vspace{0.5cm}

\nin The first term in the $k_0$-expansion leads to the integral
\bea
&&(k_0\gamma^0)\int_{p,q}\frac{N(-p)N(-q)}{D(-p)D(-q)D(p+q)}\\
&=&(k_0\gamma^0)\int_{p,q}\frac{p_0q_0-(M^2+\vec p^2)(M^2+\vec q^2)\vec p\cdot\vec q}{D(-p)D(-q)D(p+q)}\nn
&&+(k_0\gamma^0)\int_{p,q}\frac{p_0(M^2+\vec q^2)\vec q\cdot\vec\gamma-q_0(M^2+\vec q^2)\vec p\cdot\vec\gamma}{D(-p)D(-q)D(p+q)}~,\nonumber
\eea
and, because of the symmetry $p\leftrightarrow q$, the second integral vanishes. 
The following rescaling 
\be\label{rescaling}
p_0=M^3u_0~~,~~~~ q_0=M^3v_0~~,~~~~\vec p=M\vec u~~,~~~~\vec q=M\vec v ~,
\ee
together with a Wick rotation on $u_0,v_0$ finally leads to
\bea
&&(k_0\gamma^0)\int_{p,q}\frac{N(-p)N(-q)}{D(-p)D(-q)D(p+q)}\\
&=&-(k_0\gamma^0)\int\frac{d^4u_E}{(2\pi)^4}\frac{d^4v_E}{(2\pi)^4}~
\frac{u_4v_4+(1+\vec u^2)(1+\vec v^2)\vec u\cdot\vec v}{D_E(u_E)D_E(v_E)D_E(u_E+v_E)}~,
\eea
where $D_E(u_E)=u_4^2+(1+\vec u^2)^2\vec u^2$.\\
The second term in the $k_0$-expansion gives
\bea
&&-2k_0\int_{u,v}(p_0+q_0)\frac{N(-p)N(-q)N(p+q)}{D(-p)D(-q)D^2(p+q)}\\
&=&-2(k_0\gamma^0)\int_{p,q}~(p_0+q_0)^2\frac{p_0q_0-(M^2+\vec p^2)(M^2+\vec q^2)\vec p\cdot\vec q}{D(-p)D(-q)D^2(p+q)}\nn
&&-2k_0\int_{p,q}(p_0+q_0)
\frac{p_0\gamma^0~\vec q\cdot\vec\gamma~(\vec p+\vec q)\cdot\vec\gamma~(M^2+q^2)[M^2+(\vec p+\vec q)^2]
-~(p\leftrightarrow q)}{D(-p)D(-q)D^2(p+q)}~,\nonumber
\eea
where, by symmetry, the terms proportional to $\vec\gamma$ lead to a vanishing integral.
After the rescaling (\ref{rescaling}) and a 
Wick rotation, we then obtain 
\bea
&&-2k_0\int_{u,v}(p_0+q_0)\frac{N(-p)N(-q)N(p+q)}{D(-p)D(-q)D^2(p+q)}\\
&=&2(k_0\gamma^0)\int\frac{d^4u_E}{(2\pi)^4}\frac{d^4v_E}{(2\pi)^4}~(u_4+v_4)^2~
\frac{u_4v_4+(1+\vec u^2)(1+\vec v^2)\vec u\cdot\vec v}{D_E(u_E)D_E(v_E)D_E^2(u_E+v_E)}~.\nonumber
\eea
The term proportional to $k_0\gamma^0$ is then
\be\label{Y}
(k_0\gamma^0)\int\frac{d^4u_E}{(2\pi)^4}\frac{d^4v_E}{(2\pi)^4}~
\frac{u_4v_4+(1+\vec u^2)(1+\vec v^2)\vec u\cdot\vec v}{D_E(u_E)D_E(v_E)D_E(u_E+v_E)}
\left( -1+\frac{2(u_4+v_4)^2}{D_E(u_E+v_E)}\right) ~.
\ee

\vspace{0.5cm}

\nin For the first term in the $\vec k$-expansion, we use the identity
\be\label{identity}
\int_{p,q}f(p,q)~\vec p\cdot\vec k~~(\vec p+\vec q)\cdot\vec\gamma=\frac{\vec k\cdot\vec\gamma}{3}\int_{p,q}f(p,q)~\vec p\cdot(\vec p+\vec q)~,
\ee
where $f(p,q)$ depends on $(\vec p)^2,(\vec q)^2$ and $\vec p\cdot\vec q$ only.
The rescaling (\ref{rescaling}) and a Wick rotation then lead to the integral
\bea
&&-\int_{p,q}\frac{N(-p)N(-q)}{D(-p)D(-q)D(p+q)}\left[2((\vec p+\vec q)\cdot\vec k)((\vec p+\vec q)\cdot\vec\gamma)
+(M^2+(\vec p+\vec q)^2)(\vec k\cdot\vec\gamma) \right]\nn
&=&M^2(\vec k\cdot\vec\gamma)\int\frac{d^4u_E}{(2\pi)^4}\frac{d^4v_E}{(2\pi)^4}~\left( 1+\frac{5}{3}(\vec u+\vec v)^2\right) 
\frac{u_4v_4+(1+\vec u^2)(1+\vec v^2)\vec u\cdot\vec v}{D_E(u_E)D_E(v_E)D_E(u_E+v_E)}~.
\eea
The second term in the $\vec k$-expansion leads to the integral
\bea
&&2\int_{p,q}\frac{N(-p)N(-q)N(p+q)}{D(-p)D(-q)D^2(p+q)}((\vec p+\vec q)\cdot\vec k)\left[ M^2+(\vec p+\vec q)^2\right]
\left[ M^2+3(\vec p+\vec q)^2\right]\nn
&=&-2M^2\int_{u,v}\frac{u_0v_0-(1+\vec u^2)(1+\vec v^2)\vec u\cdot\vec v}{D_E(u_E)D_E(v_E)D_E^2(u_E+v_E)}
(1+(\vec u+\vec v)^2)(\vec u+\vec v)\cdot\vec\gamma\nn
&&~~~~~~~~~~~~~~\times (\vec u+\vec v)\cdot\vec k ~~[1+(\vec u+\vec v)^2][1+3(\vec u+\vec v)^2]~,
\eea
where, by symmetry, the term not proportional to $\vec\gamma$ vanishes. Using the identity (\ref{identity}), a Wick rotation then leads to
\bea
&&2\int_{p,q}\frac{N(-p)N(-q)N(p+q)}{D(-p)D(-q)D^2(p+q)}((\vec p+\vec q)\cdot\vec k)\left[ M^2+(\vec p+\vec q)^2\right]
\left[ M^2+3(\vec p+\vec q)^2\right]\nn
&=&-\frac{2}{3}M^2(\vec k\cdot\vec\gamma)\int\frac{d^4u_E}{(2\pi)^4}\frac{d^4v_E}{(2\pi)^4}~
\frac{u_4v_4+(1+\vec u^2)(1+\vec v^2)\vec u\cdot\vec v}{D_E(u_E)D_E(v_E)D_E^2(u_E+v_E)}\nn
&&~~~~~~~~~~~~~~~~~~\times(\vec u+\vec v)^2[1+(\vec u+\vec v)^2]^2[1+3(\vec u+\vec v)^2]~.
\eea
The term proportional to $(\vec k\cdot\vec\gamma)$ is then
\bea\label{Z}
&&M^2(\vec k\cdot\vec\gamma)\int\frac{d^4u_E}{(2\pi)^4}\frac{d^4v_E}{(2\pi)^4}~
\frac{u_4v_4+(1+\vec u^2)(1+\vec v^2)\vec u\cdot\vec v}{D_E(u_E)D_E(v_E)D_E(u_E+v_E)}\nn
&&~~~~~~~~~~~~~\times
\left( 1+\frac{5}{3}(\vec u+\vec v)^2-\frac{2}{3}(\vec u+\vec v)^2\frac{[1+(\vec u+\vec v)^2]^2[1+3(\vec u+\vec v)^2]}{D_E(u_E+v_E)}\right) ~.
\eea

\vspace{0.5cm}

\nin Finally, from eqs.(\ref{Y},\ref{Z}), the quantum corrections to the IR dispersion relation are determined by
\be\label{Y-Zappendix}
Y_a-Z_a=4g_a^2(3g_a^2+4g_b^2)\int\frac{d^4u_E}{(2\pi)^4}\frac{d^4v_E}{(2\pi)^4}~\mbox{Int}~,
\ee
where the integrand is
\bea
\mbox{Int}&=&\frac{1}{3}\frac{u_4v_4+(1+\vec u^2)(1+\vec v^2)\vec u\cdot\vec v}{D_E(u_E)D_E(v_E)D_E^2(u_E+v_E)}\\
&&~\times\left[ (u_4+v_4)^2\left( 6+5(\vec u+\vec v)^2\right) 
-(\vec u+\vec v)^2(1+(\vec u+\vec v)^2)^2(2+(\vec u+\vec v)^2)\right]~.\nonumber
\eea
Note that in the Lorentz-symmetric case, higher orders in $\vec u,\vec v$ are absent and $Y_a=Z_a$.
The integral (\ref{Y-Zappendix}) is evaluated as follows.

\vspace{0.5cm}

\nin We can first perform the exact integration over $u_4,v_4$, using the Feynman parametrization. 
This introduces two new variables of integration, but which lie in a compact domain of integration:
\bea
&&\frac{1}{D_E(u_E)D_E(v_E)D_E^2(u_E+v_E)}\nn
&=&6\int_0^1dx\int_0^{1-x}dy\frac{1-x-y}{[xD_E(u_E)+yD_E(v_E)+(1-x-y)D_E(u_E+v_E)]^4}~,\nonumber
\eea
We then introduce the variables $a,b,$ such that  
\be
u_4=s(a+b)~~~\mbox{and}~~~v_4=t(a-b)~,~~\mbox{with}~~s=\sqrt{1-x}~~~\mbox{and}~~~t=\sqrt{1-y}~,
\ee
to obtain
\be
\frac{du_4~dv_4}{D_E(u_E)D_E(v_E)D_E^2(u_E+v_E)}\nn
=\int_0^1dx\int_0^{1-x}dy\frac{12~da~db~st\sigma}{[2st(st+\sigma)a^2+2st(st-\sigma)b^2+D]^4}~,\nonumber
\ee
where 
\bea
D&=& x(1+u^2)^2u^2+y(1+v^2)^2v^2+\sigma(1+\Sigma)^2\Sigma\nn
\Sigma&=&(\vec u+\vec v)^2~~,~~~~\sigma=1-x-y~.
\eea
We then write, with $0\le\rho<\infty,0\le\phi<2\pi$
\be
\sqrt{2st(st+\sigma)}~a=\rho\cos\phi~~~~~\mbox{and}~~~~~\sqrt{2st(st-\sigma)}~b=\rho\sin\phi 
\ee
to obtain
\bea
&&\int du_4\int dv_4~\mbox{Int}\\
&=&2\int_0^1dx\int_0^{1-x}dy\frac{\sigma}{\sqrt{s^2t^2-\sigma^2}}\int_0^\infty\rho d\rho \int_0^{2\pi}d\phi\frac{1}{[\rho^2+D]^4}\nn
&&\times\left[\frac{\rho^2}{2}\left( \frac{\cos^2\phi}{st+\sigma}-\frac{\sin^2\phi}{st-\sigma}\right) 
+(1+u^2)(1+v^2)\vec u\cdot\vec v\right]\nn 
&&\times\left[\frac{\rho^2}{2st}\left(\frac{(s+t)^2\cos^2\phi}{st+\sigma}+\frac{(s-t)^2\sin^2\phi}{st-\sigma}\right)
 (6+5\Sigma)-\Sigma(1+\Sigma)^2(2+\Sigma) \right] \nn
&=&2\pi\int_0^1dx\int_0^{1-x}dy\frac{\sigma}{\sqrt{s^2t^2-\sigma^2}}\int_0^\infty\rho d\rho
\frac{A\rho^4+B\rho^2+C}{[\rho^2+D]^4}\nn
&=&\frac{\pi}{6}\int_0^1dx\int_0^{1-x}dy\frac{\sigma}{\sqrt{s^2t^2-\sigma^2}}\left(\frac{2A}{D}+\frac{B}{D^2}+\frac{2C}{D^3}\right) ~,
\eea
where
\bea\label{ABC}
A&=&\frac{6+5\Sigma}{4}~\frac{2s^2t^2+4\sigma^2-3\sigma(s^2+t^2)}{(s^2t^2-\sigma^2)^2}\nn
B&=&(1+u^2)(1+v^2)\vec u\cdot\vec v(6+5\Sigma)\frac{s^2+t^2-2\sigma}{s^2t^2-\sigma^2}
+\Sigma(1+\Sigma)^2(2+\Sigma)\frac{\sigma}{s^2t^2-\sigma^2}\nn
C&=&-2(1+u^2)(1+v^2)\vec u\cdot\vec v~\Sigma(1+\Sigma)^2(2+\Sigma)~.
\eea
We then define $\vec u\cdot\vec v=uv\cos\theta$ and 
\be
u=r\cos\alpha~~~,~~~v=r\sin\alpha~~~,~~~\mbox{with}~~~0\le r<\infty~~\mbox{and}~~0\le\alpha\le\pi/2~,
\ee
and the final integral is
\bea
&&F(\Lambda/M)=\int\frac{d^4u_E}{(2\pi)^4}\frac{d^4v_E}{(2\pi)^4}~\mbox{Int}\\
&=&\frac{1}{3\times2^8\pi^5}\int_0^1dx\int_0^{1-x}dy\int_0^{\Lambda/M}dr\int_0^\pi d\theta\int_0^{\pi/2}d\alpha~
\frac{\sigma~r^5\sin^2(2\alpha)\sin\theta}{\sqrt{s^2t^2-\sigma^2}}\left(\frac{2A}{D}+\frac{B}{D^2}+\frac{2C}{D^3}\right)~,\nonumber
\eea
which is quadratically divergent, as $F(z)\sim\kappa z^2$ when $z\to\infty$.
We then find via numerical integration 
\bea
&&\kappa=\lim_{z\to\infty}\left\{\frac{1}{2z}\frac{dF}{dz}\right\}\\
&=&\lim_{r\to\infty}\left\{\frac{~r^4}{3\times2^9\pi^5}\int_0^1dx\int_0^{1-x}dy\int_0^\pi d\theta\int_0^{\pi/2}d\alpha~
\frac{\sigma~\sin^2(2\alpha)\sin\theta}{\sqrt{s^2t^2-\sigma^2}}\left(\frac{2A}{D}+\frac{B}{D^2}+\frac{2C}{D^3}\right)\right\}\nn
&\simeq&-3.49\times10^{-5}~,~~~\mbox{(to a 1\% accuracy)}~.
\eea

\end{document}